\documentclass[reprint,prl, twocolumn, preprintnumbers, superscriptaddress]{revtex4}%

\usepackage{amssymb,hyperref, amsmath}
\usepackage{color}
\usepackage{graphicx}
\usepackage{multirow}

\begin{document}

\bibliographystyle{apsrev}

\preprint{JLAB-THY-11-1314}
\preprint{TCD-MATH 11-02}

\title{Isoscalar meson spectroscopy from lattice QCD}

\author{Jozef~J.~Dudek}
\email{dudek@jlab.org}
\affiliation{Jefferson Laboratory, 12000 Jefferson Avenue,  Newport News, VA 23606, USA}
\affiliation{Department of Physics, Old Dominion University, Norfolk, VA 23529, USA}

\author{Robert~G.~Edwards}
\affiliation{Jefferson Laboratory, 12000 Jefferson Avenue,  Newport News, VA 23606, USA}

\author{B\'alint~Jo\'o}
\affiliation{Jefferson Laboratory, 12000 Jefferson Avenue,  Newport News, VA 23606, USA}

\author{Michael~J.~Peardon}
\affiliation{School of Mathematics, Trinity College, Dublin 2, Ireland}

\author{David~G.~Richards}
\affiliation{Jefferson Laboratory, 12000 Jefferson Avenue,  Newport News, VA 23606, USA}

\author{Christopher~E.~Thomas}
\affiliation{Jefferson Laboratory, 12000 Jefferson Avenue,  Newport News, VA 23606, USA}

\collaboration{for the Hadron Spectrum Collaboration}

\begin{abstract}
We extract to high statistical precision an excited spectrum of single-particle isoscalar mesons using lattice QCD, including states of high spin and, for the first time, light exotic $J^{PC}$ isoscalars. 
The use of a novel quark field construction has enabled us to overcome the long-standing challenge of efficiently including quark-annihilation contributions.
Hidden-flavor mixing angles are extracted and while most states are found to be close to ideally flavor mixed, there are examples of large mixing in the pseudoscalar and axial sectors in line with experiment. 
The exotic $J^{PC}$ isoscalar states appear at a mass scale comparable to the exotic isovector states.

\end{abstract}


\maketitle 


\paragraph{Introduction:}

Mesons which have all flavor quantum numbers equal to zero, known as isoscalars, form a particularly interesting sector of QCD, the gauge-field theory of quarks and gluons. They can be constructed from quark-antiquark pairs of any flavor or even from states of pure glue, and in general the QCD eigenstates will be superpositions of these possible basis states which mix through quark annihilation. By studying the spectrum and hidden-flavor content of isoscalar mesons we can infer a phenomenology of annihilation dynamics in QCD. 

Empirically it is found that many low-lying isoscalar mesons come in identifiable pairs with a strong preference for ideal flavor mixing; prominent examples include $\omega,\phi$ where the admixture of $|s\bar{s}\rangle$ into the dominantly $\tfrac{1}{\sqrt{2}}(|u\bar{u}\rangle + |d\bar{d}\rangle)$ $\omega$ is estimated to be below $1\%$. Similarly the tensor mesons, $f_2(1270), f_2'(1525)$, where the lighter state decays almost entirely into $\pi \pi$ and the heavier into $K\bar{K}$ suggesting mixing of at most $1\%$ \cite{PDG}. 
There are notable exceptions to the dominance of ideal flavor mixing in the pseudoscalar and axial sectors. The $\eta$ and $\eta'$ mesons appear to be close to the octet and singlet eigenstates of $SU(3)_F$ which have lowest quark content $\tfrac{1}{\sqrt{6}}\left( |u\bar{u}\rangle + |d\bar{d}\rangle - 2 |s\bar{s}\rangle \right)$ and $\tfrac{1}{\sqrt{3}}\left( |u\bar{u}\rangle + |d\bar{d}\rangle + |s\bar{s}\rangle \right)$ respectively and are thus far from ideally flavor mixed. 
The scalar sector poses the greatest challenge to phenomenology
- the large number of low-lying resonances has suggested interpretations featuring light and strange $q\bar{q}$, $qq\bar{q}\bar{q}$ states, glueballs and meson-meson molecules \cite{Klempt:2007cp}.

Our aim here is to extract an isoscalar meson mass spectrum and study the hidden-flavor composition of the states using lattice QCD computations. This is a challenging undertaking for several reasons. It requires lattice gauge configurations with dynamical light and strange quarks in order that the flavor mixing appear in a unitary manner, while evaluation of the disconnected correlator contributions that distinguish isoscalars from isovectors is computationally expensive and the signals obtained typically diminish into noise at small Euclidean times. These problems have limited calculations to a few $J^{PC}$ with typically low statistical precision \cite{Christ:2010dd, McNeile:2009mx, McNeile:2001cr}. Glueball studies, which produce exceptionally clean spectra in the quark-less Yang-Mills case \cite{Morningstar:1999rf}, become challenging in QCD through strong coupling to the quark sector  \cite{Richards:2010ck}.

In this letter we present the isoscalar meson spectrum for a single choice of light and strange quark masses, lattice spacing and volume across multiple $J^{PC}$, excluding the $0^{++}$ case which is of sufficient interest to justify a separate publication of its own. This extends the work reported in \cite{Dudek:2009qf,Dudek:2010wm} for isovector and kaonic mesons, taking advantage of many of the techniques developed therein.

\paragraph{Lattice technology:} 
We compute the spectrum of isoscalar mesons using a basis of operators of the form 
${\cal O}_{A}^\ell(t) = \tfrac{1}{\sqrt{2}} \left( \bar{u} \boldsymbol{\Gamma}^{A}_t u + \bar{d}\boldsymbol{\Gamma}^A_td \right)$, ${\cal O}_A^s(t) = \bar{s}\boldsymbol{\Gamma}^A_t s $, where $u$, $d$, and $s$ are the up, down and strange quark fields, and the $\boldsymbol{\Gamma}^A_t$ are operators acting in space, color, and Dirac spin space~\cite{Peardon:2009gh} on a time slice, $t$. We combine these operators to construct two-point correlators of the form $C^{q'q}_{AB}(t',t)= \langle 0 | {\cal O}^{q'}_A(t') {\cal O}^{q\dag}_B(t)| 0 \rangle$ which, after integration of the quark fields, can be composed from 
\emph{connected} components, diagonal in quark flavor,
$${\cal C}^{q'q}_{AB}(t',t) = \delta_{qq'} \mathrm{Tr}\big[ \Phi^A(t') \tau_{q'}(t',t) \Phi^B(t) \tau_{q}(t,t') \big], \nonumber$$
and \emph{disconnected} components that can mix flavor,
\begin{equation}
{\cal D}^{q'q}_{AB}(t',t) = \mathrm{Tr}\big[ \Phi^A(t') \tau_{q'}(t',t') \big]\; \mathrm{Tr}\big[ \Phi^B(t) \tau_{q}(t,t) \big]. \nonumber
\label{eq:matrix}
\end{equation}

The quark fields in ${\cal O}$ are acted upon by a ``distillation'' smearing operator that emphasizes the low momentum quark and gluon modes that dominate low mass hadrons~\cite{Peardon:2009gh, Dudek:2010wm}. This smearing can be factorized allowing the ``perambulators", $\tau_q$, and the zero momentum projected operators, $\Phi$, to be constructed as matrices in distillation space: $\tau_q(t',t) = V^{\dag}_{t'}M^{-1}_q(t',t) V_t$ and $\Phi^A(t)=V^{\dag}_t \boldsymbol{\Gamma}^A_t V_t$, where $V_t$ are the eigenvectors of the gauge-covariant three-dimensional Laplacian operator on a time-slice $t$, and $M_q$ is the lattice representation of the Dirac operator for quark flavor $q$. 
The resulting traces are then over the set of eigenvectors which is much smaller than the full lattice space - this allows for an efficient computation of the disconnected terms, $\cal D$, that is not possible in the traditional ``point-to-all" method.

We consider the limit where the $u$ and $d$ quark masses are set equal, and are replaced by a ``light'' quark, $\ell$. For a particular combination of creation and annihilation operators ($A,B$), the flavor subspace in the correlation matrix is then
\begin{equation}
C = \begin{pmatrix}
-{\cal C}^{\ell \ell} + 2\, {\cal D}^{\ell\ell} &  \sqrt{2}\, {\cal D}^{\ell s} \\ 
 \sqrt{2}\, {\cal D}^{s \ell} & -{\cal C}^{ss} +  {\cal D}^{ss} 
\end{pmatrix}.\nonumber
\end{equation} 
The signals for $\cal D$ and $\cal C$ are typically short in physical units. To help resolve this, we use anisotropic lattices with three dynamical flavors of Clover fermions~\cite{Edwards:2008ja,Lin:2008pr} of size $16^3\times 128$, spatial lattice spacing $a_s \sim 0.12 \,\mathrm{fm}$, and temporal spacing, $a_t^{-1} \sim 5.6 \, \mathrm{GeV}$. This fine temporal lattice spacing has proven useful in determining the spectrum of isovector mesons~\cite{Dudek:2009qf,Dudek:2010wm} and baryons~\cite{Bulava:2010yg}.

\begin{figure}
 \centering
\includegraphics[width=.49 \textwidth, bb=10 10 380 220]{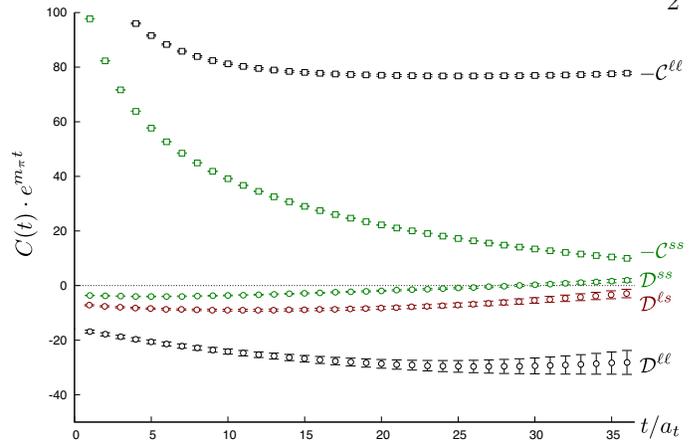} 
\caption{Time-source averaged connected and disconnected correlators with $\boldsymbol{\Gamma}_A = \boldsymbol{\Gamma}_B = \gamma^5 $.\label{corrs}}
\end{figure}

The disconnected terms $\cal D$ require perambulators $\tau_q(t,t)$ with sources on every time-slice. 
With $\tau_q(t',t_0)$ from all timeslice sources, $t_0$, in hand, we can maximize the signal above noise by averaging:
${\cal D}^{q'q}(t) = \frac{1}{L_t} \sum_{t_0}^{L_t} {\cal D}^{q'q}(t+t_0, t_0)$, 
 ${\cal C}^{qq}(t) = \frac{1}{N_t} \sum_{t_0}^{N_t} {\cal C}^{qq}(t+t_0, t_0)$, 
where the disconnected term is averaged over all $L_t=128$ timeslices, while the connected term needs far less averaging to achieve a comparable level of statistical fluctuation - in this study we used $N_t=32$. Since the eigenvectors $V_t$ sample the full three-dimensional space and the inversions are from all time-slices, the propagation from a source is from every point of the four-dimensional lattice, resulting in a significant decrease in the noise of the correlator compared to ``point-to-all'' methods. 
In all, 479 configurations (separated by 20 trajectories) were used with 64 distillation vectors. 
Graphics Processing Units~\cite{Clark:2009wm,Babich:2010mu} allow us to evaluate the 31-million Dirac propagator inversions required~\footnote{Autocorrelations are found to be small - binning measurements by 10 produces no significant change in the final spectra.}. An example of the correlator quality is shown in Figure \ref{corrs} for the case of the distilled $\gamma_5$ operator. The disconnected contributions are resolved for time-separations beyond $1$ fm.

\paragraph{Extraction of the spectrum:} Our approach is to form a matrix of correlators with elements $C_{AB}^{qq'}(t)$ that is then diagonalized in the (operator $\otimes$ flavor) space to obtain a spectrum which best describes the correlators in the variational sense \cite{Dudek:2010wm}. This yields both estimates of the mass spectrum, $\{m_{\mathfrak{n}}\}$, and the overlap of states onto the operators used to interpolate from the vacuum, $Z^{\mathfrak{n}}_{q,A} \equiv \langle \mathfrak{n} |{\cal O}^{q\dag}_A(0)|0\rangle $. The overlaps can then be studied to explore the hidden-flavor composition of the states. 
We use an operator set founded upon the derivative-based constructions for $\boldsymbol{\Gamma}$ described in \cite{Dudek:2010wm}, doubled in size by forming operators from both light-quark ($q=\ell$) and strange-quark ($q=s$) fields.

\begin{figure}
 \centering
\includegraphics[width=.49 \textwidth, bb=15 0 363 180]{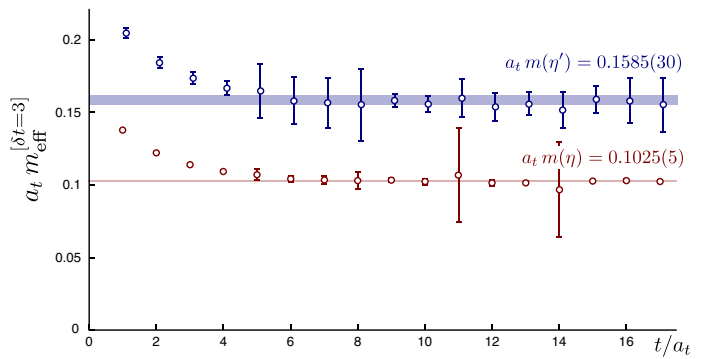} 
\caption{Effective mass of lowest two variational principal correlators in the pseudoscalar channel, $m_{\mathrm{eff}}^{[\delta t]} = -\frac{1}{\delta t} \ln \left[ \frac{\lambda(t + \delta t)}{\lambda(t)} \right]$, with $\delta t = 3$. The non-monotonic behavior of the statistical error is an artifact of our implementation of the variational analysis method. The horizontal bands indicate the mass result from fits to the principal correlators.
\label{meff}}
\end{figure}

The variational analysis is performed independently in each cubic-group irreducible representation, $\Lambda^{PC}$, resulting in spectra that, as in \cite{Dudek:2010wm}, conform to the mass degeneracy and overlap patterns expected of \emph{single-particle} states of good $J^{PC}$ subduced into the relevant $\Lambda^{PC}$. 
We find that while isoscalar signals are typically noisier than the corresponding isovector correlators, they are not excessively so and a detailed spectrum comparable to that presented in \cite{Dudek:2010wm} for isovector states follows.
As an example of the quality of spectrum signals observed, in Figure \ref{meff} we show the effective mass of the principal correlators for the lightest two states with $0^{-+}$ quantum numbers which we identify with the $\eta$ and $\eta'$ mesons. 

\begin{figure}
 \centering
\includegraphics[width=.45 \textwidth, bb=15 0 357 240]{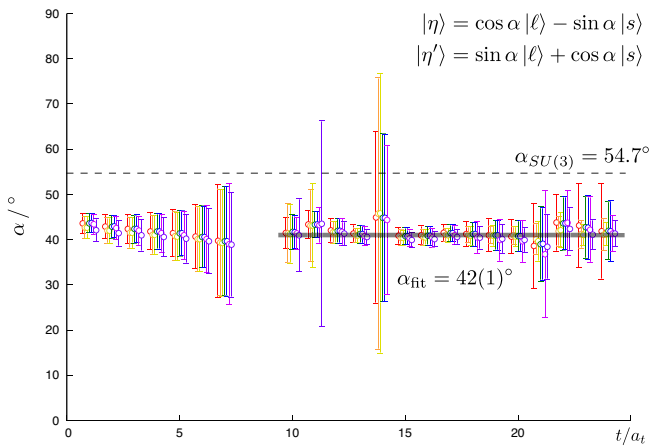} 
\caption{Effective mixing angle, $\alpha_A(t)$, as a function of Euclidean time for the lightest two pseudoscalar states. Colors indicate seven different operator constructions, $\boldsymbol{\Gamma}_A$, and the grey band a constant fit. 
\label{mix}}
\end{figure}

Within a given $J^{PC}$ sector of the extracted spectrum, we find that states occur in pairs with overlaps that suggest they are admixtures of the light-quark, strange-quark basis states. We parameterize this admixture by introducing a mixing angle, assuming the pair of states ($\mathfrak{a}$, $\mathfrak{b}$) are orthogonal combinations of just two light-strange basis states: 
\begin{align}
\big| \mathfrak{a} \big\rangle &= \cos \alpha \big| \ell \big\rangle - \sin \alpha \big| s\big\rangle \nonumber\\
\big| \mathfrak{b} \big\rangle &=  \sin \alpha \big| \ell \big\rangle + \cos \alpha \big| s\big\rangle \nonumber
\end{align}
where $|\ell \rangle \equiv \tfrac{1}{\sqrt{2}}\left( |u\bar{u}\rangle + |d\bar{d}\rangle \right)$, $|s\rangle \equiv |s\bar{s}\rangle$ and where $\mathfrak{a}$ is the lighter of the two states.
The mixing angle is determined~\footnote{Since each state can be arbitrarily rephased without changing the correlator matrix, the sign of $\alpha$ cannot be determined in this way. The angle $\alpha$ is related to the octet-singlet mixing angle\cite{PDG}, $\theta$ by $\theta = \alpha - 54.74^\circ$.} for each operator, $\boldsymbol{\Gamma}^A$,  from 
$\alpha_A = \tan^{-1}\left[\sqrt{ - \frac{Z^{\mathfrak{b}}_{\ell,A} \, Z^{\mathfrak{a}}_{s,A} }{Z^{\mathfrak{a}}_{\ell,A} \, Z^{\mathfrak{b}}_{s,A}} }\right]$, 
where the overlaps $Z$ are determined on each timeslice by the variational diagonalization and where 
we're assuming that e.g. $\langle \ell | {\cal O}_A^{s\dag} | 0 \rangle = 0$ to a good approximation. In general we find a good agreement between determinations from the different operators forming our variational basis - see Figure \ref{mix} for the example of the $\eta,\eta'$ states.

The quark-bilinear operator basis thus far described is imperfect for the study of isoscalar mesons since it lacks constructions likely to have a large overlap onto glueball basis states. 
Such operators can be constructed from gauge-invariant link paths on the lattice, and it is known that in the quark-less Yang-Mills $SU(3)$-color theory there is a spectrum of bound states, the lightest of which fall within the mass range considered in our study~\cite{Morningstar:1999rf}. 
We included operators of this type in the basis but it was found that correlators featuring them had signals which degraded into noise at rather short Euclidean times before any clear state signals could be extracted. Any resolution of such states will require operators that plateau at shorter Euclidean times as well as more gauge configurations. 
The presence of torelons, states of gluonic flux wrapped around the periodic spatial volume, while anticipated in the studied mass region, are not observed in the spectrum.

\begin{figure*}
 \centering
\includegraphics[width=0.95\textwidth, bb=25 20 750 420]{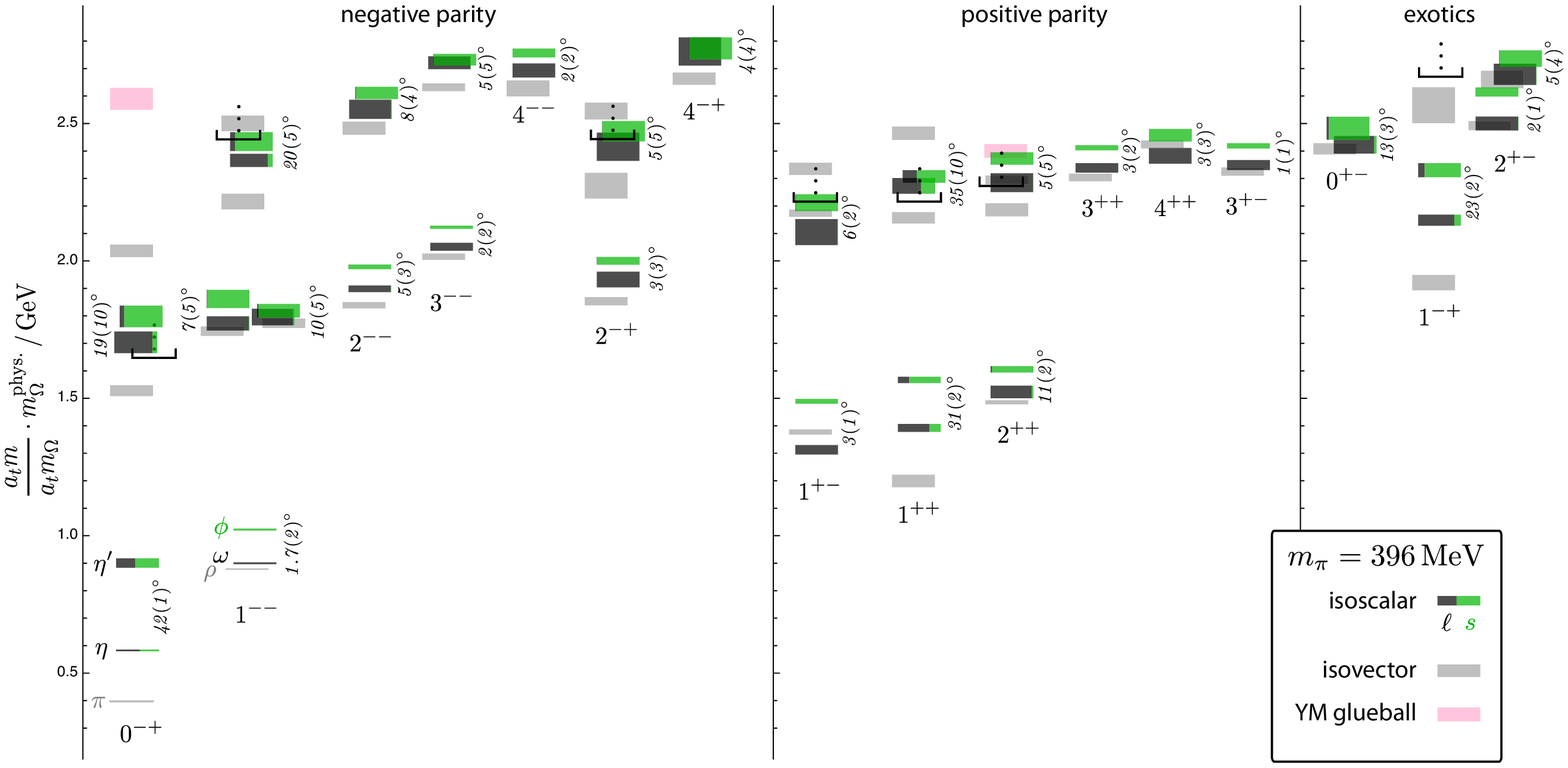} 
\caption{Isoscalar meson spectrum labeled by $J^{PC}$. The box height indicates the one sigma statistical uncertainty above and below the central value. The light-strange content of each state ($\cos^2 \alpha, \sin^2 \alpha$) is given by the fraction of (black, green) and the mixing angle for identified pairs is also shown. Horizontal square braces with ellipses indicate that additional states were extracted in this $J^{PC}$ but were not robust.  Grey boxes indicate the positions of isovector meson states extracted on the same lattice (taken from \cite{Dudek:2010wm}). The mass scale is set using the procedure outlined in \cite{Dudek:2010wm, Lin:2008pr}. Pink boxes indicate the position of glueballs in the quark-less Yang-Mills theory \cite{Morningstar:1999rf}.}
\label{spectrum}
\end{figure*}

In Figure \ref{spectrum} we present the main result of this study, the spectrum of isoscalar mesons with $m_\pi = 396$ MeV on a $\sim (2\,\mathrm{fm})^3$ lattice identified by $J^{PC}$ quantum numbers. The degree of light-strange mixing is indicated by the black-green coloring and by the displayed mixing angle $\alpha$. Superimposed on the plot is the spectrum of isovector mesons on the same lattice, taken from reference \cite{Dudek:2010wm}, and for comparison the spectrum of glueballs in the quark-less Yang-Mills theory taken from \cite{Morningstar:1999rf}.

\paragraph{Phenomenology:} The $m_\pi \sim 400$ MeV spectrum results presented in Figure \ref{spectrum} suggest a phenomenology of isoscalar mesons that is in quite good agreement with experiment. 
We reproduce the pattern of $\eta$, $\eta'$ observed in experiment and suggested by the axial anomaly of QCD, where the lighter $\eta$ meson is dominantly an $SU(3)_F$ octet and the significantly heavier $\eta'$ is dominantly $SU(3)_F$ singlet. In fact our estimate of the light-strange mixing angle, $42(1)^\circ$, deviates somewhat from the exact $SU(3)_F$ expectation of $54.7^\circ$, in line with phenomenological determinations (e.g.~\cite{Mathieu:2009sg,Thomas:2007uy,Escribano:2007cd} and references therein). 

The level of statistical precision in the disconnected contribution to the vector channel is such that we can determine the $\omega -\rho$ mass splitting to be $21(5)$ MeV. The very small $1.7(2)^\circ$ $\omega/\phi$ mixing angle is comparable to the $3.2(1)^\circ$ inferred from analysis of radiative decay rates of these states to pseudoscalar mesons~\cite{Escribano:2007cd}.

We observe that the bulk of states are close to ideally flavor mixed, and we find that the majority of light isoscalar states are heavier than their corresponding isovector cousins even in the cases of negligible admixture of heavier $s\bar{s}$ components. This is of course possible through the disconnected contribution ${\cal D}_{\ell \ell}$ to the correlator and hence the effect of ``annihilation dynamics".
An exception to this observation is the $1^{+-}$ channel, where the lightest $h_1$, with negligible $s\bar{s}$ component, is lighter than the isovector $b_1$, in line with experiment, although given the large hadronic decay width of the $h_1$ this may not be significant. Apart from the previously discussed pseudoscalars, the only other case of large mixing angles in a conventional $J^{PC}$ channel are the axial ($1^{++}$) states. For the lightest two states we find a mixing angle of $31(2)^\circ$ which is in reasonable agreement with the phenomenologically determined~\footnote{Using the radiative transitions methodology of \cite{Close:1997nm} applied to current PDG \cite{PDG} branching fraction values.} $21(5)^\circ$ between the $f_1(1285)$ and the $f_1(1420)$.

The rightmost section of Figure \ref{spectrum} shows our determination of the spectrum of \emph {exotic} $J^{PC}$ isoscalar mesons. This is the first computation of such a spectrum and indicates that isoscalar exotics appear at a comparable mass scale to the exotic isovectors, and as in \cite{Dudek:2010wm}, the pattern of operator overlaps suggests that these states are hybrid mesons. While the $0^{+-}$ and $2^{+-}$ states are rather close to ideally flavor mixed, there is considerable mixing in the $1^{-+}$ channel. 

\paragraph{Outlook:} 
In summary, we find that lattice QCD can reproduce the pattern of single-particle light isoscalar states observed in nature, and the determined flavor mixings are in reasonable agreement with phenomenology. Beyond their role in the isoscalar spectrum, annihilation dynamics feature prominently in hadron decays. Combining the methods developed in this letter with finite-volume techniques for the extraction of phase shifts~\cite{Dudek:2010ew} and stochastic methods~\cite{Foley:2010vv}, future work will concentrate on the determination of hadronic resonances within QCD.


\paragraph{Acknowledgments:}
We thank our colleagues within the Hadron Spectrum Collaboration. 
{\tt Chroma}~\cite{Edwards:2004sx} and {\tt QUDA}~\cite{Clark:2009wm,Babich:2010mu} were used to perform this work 
on clusters at Jefferson Laboratory under the USQCD 
Initiative and the LQCD ARRA project. Support is from Jefferson Science Associates, LLC under U.S. DOE Contract No.
{\small{\sc DE-AC05-06OR23177}} 
and Science Foundation Ireland under research grant {\small{\sc 07/RFP/PHYF168}}.


\bibliography{isoscalar}

\end{document}